\documentclass[twocolumn]{emulateapj}
\usepackage{graphicx}

\catcode`\@=11
\def\gsim{\ifmmode{\,\mathrel{\mathpalette\@versim>\,}}
    \else{$\,\mathrel{\mathpalette\@versim>}\,$}\fi}
\def\lsim{\ifmmode{\,\mathrel{\mathpalette\@versim<\,}}
    \else{$\,\mathrel{\mathpalette\@versim<}\,$}\fi}
\def\@versim#1#2{\lower 2.9truept \vbox{\baselineskip 0pt \lineskip
    0.5truept \ialign{$\m@th#1\hfil##\hfil$\crcr#2\crcr\sim\crcr}}}
\catcode`\@=12
\newcommand{\gv}{{\bf g}}

\newcommand{\xv}{{\bf x}}
\newcommand{\Sv}{{\bf S}}

\newcommand{\Mstar}{{M_*}}

\newcommand{\rstar}{r_*}
\newcommand{\rt}{r_{\rm t}}
\newcommand{\tstar}{t_*}
\newcommand{\az}{a_0}

\newcommand{\phiN}{\phi_{\rm N}}
\newcommand{\phistar}{\phi_*}
\def\sglos{\sigma_{\rm los}}

\newcommand{\uvec}{{\bf u}}
\newcommand{\diverg}{\nabla\cdot}

\newcommand{\vstarn}{v_{\rm *n}}

\newcommand{\tdyn}{t_{\rm dyn}}
\newcommand{\tstard}{t_{\rm *d}}
\newcommand{\tstarn}{t_{\rm *n}}
\def\Ie{I_{\rm e}}

\def\Re{R_{\rm e}}

\def\Dt{\Delta t}
\def\sae{a_{\rm e}}
\def\sbe{b_{\rm e}}
\def\sgz{\sigma_0}
\def\sgv{\sigma_{\rm V}}
\newcommand{\vstard}{v_{\rm *d}}
\newcommand{\Estard}{E_{\rm *d}}
\newcommand{\Estarn}{E_{\rm *n}}

\newcommand{\be}{\begin{equation}}
\newcommand{\ee}{\end{equation}}

\def\dxcube{d^3{\bf x}}

\def\Nr{N_{r}}

\def\Nth{N_{\vartheta}}
\def\Nph{N_{\varphi}}

\def\trho{\tilde{\rho}}

\def\Tr{{\rm Tr}\,}
\def\tphi{\tilde{\phi}}
\def\txv{\tilde{\xv}}
\def\tnabla{\tilde{\nabla}}

\def\vr{v_r}

\newcommand{\hv}{{\bf h}}

\def\Myr{\rm\,Myr}
\def\Gyr{\rm\,Gyr}
\def\kpc{\rm\,kpc}
\def\Msun{M_{\odot}}
\def\rhalf{r_{\rm h}}
\def\rc{r_{\rm c}}
\def\Dcal{{\cal D}}

\def\Dprime{{\rm D^{\prime}}}
\def\Mprime{{\rm M^{\prime}}}
\def\Nprime{{\rm N^{\prime}}}

\begin{document}
\slugcomment{Accepted, January 15, 2007}
\shorttitle{Dissipationless collapses in MOND}
\shortauthors{Nipoti, Londrillo \& Ciotti}
\title{Dissipationless collapses in MOND}
\author{Carlo Nipoti}
\affil{Dept. of Astronomy, University of Bologna, I-40127 Bologna, Italy}
\email{carlo.nipoti@unibo.it}
\author{Pasquale Londrillo}
\affil{INAF-Bologna Astronomical Observatory, I-40127 Bologna, Italy}
\author{Luca Ciotti}
\affil{Dept. of Astronomy, University of Bologna, I-40127 Bologna, Italy}

\begin{abstract}
Dissipationless collapses in Modified Newtonian Dynamics (MOND) are
studied by using a new particle-mesh N-body code based on our
numerical MOND potential solver.  We found that low surface-density
end-products have shallower inner density profile, flatter radial
velocity-dispersion profile, and more radially anisotropic orbital
distribution than high surface-density end-products.  The projected
density profiles of the final virialized systems are well described by
Sersic profiles with index $m \lsim 4$, down to $m\sim 2$ for a
deep-MOND collapse. Consistently with observations of elliptical
galaxies, the MOND end-products, if interpreted in the context of
Newtonian gravity, would appear to have little or no dark matter
within the effective radius.  However, we found impossible (under the
assumption of constant mass-to-light ratio) to simultaneously place
the resulting systems on the observed Kormendy, Faber-Jackson and
Fundamental Plane relations of elliptical galaxies.  Finally, the
simulations provide strong evidence that phase mixing is less
effective in MOND than in Newtonian gravity.
\end{abstract}


\section{Introduction}

In Bekenstein \& Milgrom's (1984, hereafter BM) Lagrangian formulation
of Milgrom's (1983) Modified Newtonian Dynamics (MOND), the Poisson
equation 
\begin{equation}
\nabla^2\phiN=4\pi G\rho
\label{eqPoisson}
\end{equation}
for the Newtonian gravitational potential $\phiN$ is replaced by the
field equation
\begin{equation}
\nabla\cdot\left[\mu\left({\Vert\nabla\phi\Vert\over\az}\right)
                 \nabla\phi\right] = 4\pi G \rho,
\label{eqMOND}
\end{equation}
where $\az\simeq 1.2 \times 10^{-10} {\rm m\, s^{-2}}$ is a
characteristic acceleration, $\Vert ...\Vert$ is the standard
Euclidean norm, $\phi$ is the MOND gravitational potential produced by
the density distribution $\rho$, and in finite mass systems
$\nabla\phi\to 0$ for $\Vert\xv\Vert\to\infty$.  The MOND
gravitational field $\gv$ experienced by a test particle is
\begin{equation}
\gv=-\nabla\phi,
\label{eqgv}
\end{equation}
and the function $\mu$ is such that
\begin{equation}
\mu(y)\sim\cases{y&for $y\ll 1$,\cr 1&for $y\gg 1$;}
\label{eqmulim}
\end{equation}
throughout this paper we use
\begin{equation}
\mu (y)={y\over\sqrt{1+y^2}}.
\label{eqmu}
\end{equation}
In the so-called `deep MOND regime' (hereafter dMOND), describing
low-acceleration systems ($\Vert\nabla\phi\Vert \ll\az$), $\mu(y)=y$
and so equation (\ref{eqMOND}) simplifies to 
\begin{equation}
\nabla\cdot\left({\Vert\nabla\phi\Vert}\nabla\phi\right) = 4\pi G \az \rho.
\label{eqdMOND}
\end{equation}
The source term in equation~(\ref{eqMOND}) can be eliminated by using
equation~(\ref{eqPoisson}), giving
\begin{equation}
\mu\left(\Vert\nabla\phi\Vert\over\az\right)\nabla\phi=\nabla\phiN+\Sv,
\label{eqcurl}
\end{equation}
where $\Sv={\rm curl\,}\hv$ is a solenoidal field dependent on $\rho$
and in general different from zero. When $\Sv=0$
equation~(\ref{eqcurl}) reduces to Milgrom's~(1983) formulation and
can be solved explicitly. Such reduction is possible for
configurations with spherical, cylindrical or planar symmetry, which
are special cases of a more general family of stratifications (BM;
Brada \& Milgrom~1995). Though the solenoidal field $\Sv$ has been
shown to be small for some configurations (Brada \& Milgrom~1995;
Ciotti, Londrillo \& Nipoti~2006, hereafter CLN), neglecting it when
simulating time-dependent dynamical processes has dramatic effects
such as non-conservation of total linear momentum (e.g. Felten~1984;
see also Section~\ref{secic}).
 
Nowadays several astronomical observational data appear consistent
with the MOND hypothesis (see, e.g., Milgrom~2002; Sanders \&
McGaugh~2002). In addition, Bekenstein~(2004) recently proposed a
relativistic version of MOND (Tensor-Vector-Scalar theory, TeVeS),
making it an interesting alternative to the cold dark matter
paradigm. However, dynamical processes in MOND have been investigated
very little so far, mainly due to difficulties posed by the
non-linearity of equation~(\ref{eqMOND}). Here we recall the
spherically symmetric simulations (in which $\Sv=0$) of gaseous
collapse in MOND by Stachniewicz \& Kutschera~(2005) and Nusser \&
Pointecouteau~(2006).  The only genuine three-dimensional MOND N-body
simulations (in which equation~[\ref{eqMOND}] is solved exactly) are
those by Brada \& Milgrom~(1999, 2000), who studied the stability of
disk galaxies and the external field effect, 
and those of Tiret \& Combes (2007). Other attempts to study
MOND dynamical processes have been conducted using three-dimensional
N-body codes by arbitrarily setting $\Sv=0$: Christodoulou~(1991)
investigated disk stability, while Nusser~(2002) and Knebe \&
Gibson~(2004) explored cosmological structure
formation\footnote{Cosmological N-body simulations in the context of a
relativistic MOND theory such as TeVeS have not been performed so
far.}.

In this paper we present results of N-body simulations of
dissipationless collapse in MOND.  The simulations were performed with
an original three-dimensional particle-mesh N-body code, based on the
numerical MOND potential solver presented in CLN, which solves
equation~(\ref{eqMOND}) exactly.  These numerical experiments are
interesting both from a purely dynamical point of view, allowing for
the first time to explore the relaxation processes in MOND, and in the
context of elliptical galaxy formation. In fact, the ability of
dissipationless collapse at producing systems strikingly similar to
real ellipticals is a remarkable success of Newtonian dynamics (e.g.,
van Albada~1982; Aguilar \& Merritt~1990; Londrillo, Messina \&
Stiavelli~1991; Udry~1993; Trenti, Bertin \& van Albada~2005; Nipoti,
Londrillo, \& Ciotti~2006, hereafter NLC06), while there have been no
indications so far that MOND can work as well in this respect. Here we
study the structural and kinematical properties of the end-products of
MOND simulations, and we compare them with the observed scaling
relations of elliptical galaxies: the Faber--Jackson (FJ) relation
(Faber \& Jackson 1976), the Kormendy (1977) relation, and the
Fundamental Plane (FP) relation (Djorgovski \& Davis 1987, Dressler et
al. 1987).

The paper is organized as follows.  The main features of the new
N-body code are presented in Section~\ref{seccod}, while
Section~\ref{secsim} describes the set-up and the analysis of the
numerical simulations. The results are presented in
Section~\ref{secres} and discussed in Section~\ref{secdis}.

\section{The N-body code}
\label{seccod}

While most N-body codes for simulations in Newtonian gravity are based
on the gridless multipole expansion treecode scheme (Barnes \& Hut
1986; see also Dehnen~2002), the non-linearity of the MOND field
equation~(\ref{eqMOND}) forces one to resort to other methods, such as
the particle-mesh technique (see Hockney \& Eastwood~1988).  In this
approach, particles are moved under the action of a gravitational
field which is computed on a grid, with particle-mesh interpolation
providing the link between the two representations.  In our MOND
particle-mesh N-body code, we adopt a spherical grid of coordinates
($r$, $\vartheta$, $\varphi$), made of $\Nr\times\Nth\times\Nph$
points, on which the MOND field equation is solved as in
CLN. Particle-mesh interpolations are obtained with a quadratic spline
in each coordinate, while time stepping is given by a classical
leap-frog scheme (Hockney \& Eastwood~1988).  The time-step $\Dt$ is
the same for all particles and is allowed to vary adaptively in
time. In particular, according to the stability criterion for the
leap-frog time integration, we adopt $\Dt=\eta/\sqrt{\max{|\nabla^2
\phi|}}$, where $\eta\lsim0.3$ is a dimensionless parameter. We found
that $\eta=0.1$ assures good conservation of the total energy in the
Newtonian cases (see Section~\ref{secic}). In the present version of
the code, all the computations on the particles and the particle-mesh
interpolations can be split among different processors, while the
computations relative to the potential solver are not performed in
parallel. The solution of equation~(\ref{eqMOND}) over the grid is
then the bottleneck of the simulations: however, the iterative
procedure on which the potential solver is based (see CLN) allows to
adopt as seed solution at each time step the potential previously
determined.

The MOND potential solver can also solve the Poisson equation
(obtained by imposing $\mu=1$ in equation~\ref{eqMOND}), so Newtonian
simulations can be run with the same code.  We exploited this property
to test the code by running several Newtonian simulations of both
equilibrium distributions and collapses, comparing the results with
those of simulations (starting from the same initial conditions)
performed with the FVFPS treecode (Londrillo, Nipoti \& Ciotti~2003;
Nipoti, Londrillo \& Ciotti~2003).  One of these tests is described in
Section~\ref{seckin}.  

We also verified that the code reproduces the Newtonian and MOND
conservation laws (see Section~\ref{secic}): note that the
conservation laws in MOND present some peculiarities with respect to
the Newtonian case, so we give here a brief discussion of the
subject. As already stressed by BM, equation~(\ref{eqMOND}) is
obtained from a variational principle applied to a Lagrangian with all
the required symmetries, so energy, linear and angular momentum are
conserved. Unfortunately, as also shown by BM, the total energy
diverges even for finite mass systems, thus posing a computational
challenge to code validation. We solved this problem by checking the
volume-limited energy balance equation
\begin{eqnarray} 
\label{eqbaltext}
{d \over d t}\int_{V_0}\left[k+\rho\phi+{\az^2\over 8 \pi
G}\mathcal{F}\left({||\nabla\phi||\over
    \az}\right)\right]\dxcube=\nonumber\\{1\over 4\pi G}\int_{\partial
  V_0}\mu{\partial \phi \over \partial t} <\nabla \phi,\hat{\bf n}> d a,
\end{eqnarray} 
which is derived in Appendix \ref{appetot}. In equation
(\ref{eqbaltext}) $V_0$ is an arbitrary (but fixed) volume enclosing
all the system mass, $k$ is the kinetic energy per unit volume, and
\begin{equation}
\mathcal{F}(y)\equiv2\int_{y_0}^{y} \mu(\xi) \xi d \xi,
\label{eqeffe}
\end{equation}
where $y_0$ is an arbitrary constant; note that only finite quantities
are involved.  Another important relation between global quantities
for a system at equilibrium (in MOND as in Newtonian gravity) is the
virial theorem
\begin{equation}\label{eqvirtheo}
2 K + W=0,
\end{equation}
where $K$ is the total kinetic energy and $W=\Tr W_{ij}$ is the trace
of the Chandrasekhar potential energy tensor
\begin{equation}\label{eqwij}
W_{ij}\equiv-\int \rho(\xv) x_i {\partial \phi(\xv) \over \partial x_j} \dxcube
\end{equation}
(e.g., Binney \& Tremaine 1987). Note that in MOND $K+W$ is {\it not}
the total energy, and is not conserved. However, {\it $W$ is conserved
in the limit of dMOND}, being $W=-(2/3)\sqrt{G\az M_*^3}$ for {\it
all} systems of finite total mass $\Mstar$ (see Appendix~\ref{appw}
for the proof).  As a consequence, in dMOND the virial theorem writes
simply $\sgv^4=4G\Mstar\az/9$, where $\sgv \equiv \sqrt{2 K / \Mstar}$
is the system virial velocity dispersion (this relation was proved for
dMOND spherical systems by Gerhard \& Spergel~1992; see also
Milgrom~1984).  In our simulations we also tested that equation
(\ref{eqvirtheo}) is satisfied at equilibrium, and that $W$ is
conserved in the dMOND case (see Sections \ref{secic} and
\ref{secres}).

\section{Numerical simulations}
\label{secsim}

\begin{table}
\centering
\caption{Time, velocity, and energy units for Newtonian and MOND
  (subscript n), and dMOND (subscript d) N-body simulations.}
  \begin{tabular}{ll}
\\
\hline
$\tstarn=\rstar^{3/2} (G \Mstar)^{-1/2}$  & $\tstard=\rstar (G \Mstar\az)^{-1/4}$ \\
$\vstarn=(G \Mstar)^{1/2}\rstar^{-1/2}$ & $\vstard=(G \Mstar \az)^{1/4}$ \\
$\Estarn=G \Mstar^2\rstar^{-1}$  & $\Estard=(G\az)^{1/2}\Mstar^{3/2}$ \\
\hline
\end{tabular}
\end{table}
The choice of appropriate scaling physical units is an important
aspect of N-body simulations. This is especially true in the present
case, in which we want to compare MOND and Newtonian simulations
having the same initial conditions.  As well known, due to the
scale-free nature of Newtonian gravity, a Newtonian $N$-body
simulation starting from a given initial condition describes in
practice $\infty^2$ systems of arbitrary mass and size. Each of them
is obtained by assigning specific values to the length and mass units,
$\rstar$ and $\Mstar$, in which the initial conditions are expressed.
Also dMOND gravity is scale free, because $\az$ appears only as a
multiplicative factor in equation~(\ref{eqdMOND}), and so a simulation
in dMOND gravity represents systems with arbitrary mass and size
(though in principle the results apply only to systems with
accelerations much smaller than $\az$).  MOND simulations can also be
rescaled, but, due to the presence of the characteristic acceleration
$\az$ in the non-linear function $\mu$, each simulation describes only
$\infty^1$ systems, because $\rstar$ and $\Mstar$ cannot be chosen
independently of each other.
 
On the basis of the above discussion, we fix the physical units as
follows (see Appendix~\ref{appscal} for a detailed description of the
scaling procedure). Let the initial density distribution be
characterized by a total mass $\Mstar$ and a characteristic radius
$\rstar$. We rescale the field equations so that the dimensionless
source term is the same in Newtonian, MOND and dMOND simulations. We
also require that the Second Law of Dynamics, when cast in
dimensionless form, is independent of the specific force law
considered, and this leads to fix the time unit.  As a result,
Newtonian and MOND simulations have the same time unit
$\tstarn=\rstar^{3/2} (G \Mstar)^{-1/2}$, while the natural time unit
in dMOND simulations is $\tstard=\rstar (G \Mstar \az)^{-1/4}$. Note
that MOND simulations are characterized by the dimensionless parameter
$\kappa=G \Mstar /\rstar^2\az$, and scaling of a specific simulation
is allowed provided the value of $\kappa$ is maintained constant.  So,
simulations with lower $\kappa$ values describe lower surface-density,
weaker acceleration systems; dMOND simulations represent the limit
case $\kappa \ll 1$, while Newtonian ones describe the regime with
$\kappa \gg 1$. With the time units fixed, the corresponding velocity
and energy units are $\vstarn\equiv\rstar/\tstarn$,
$\vstard\equiv\rstar/\tstard$, $\Estarn=\Mstar\vstarn^2$, and
$\Estard=\Mstar\vstard^2$ (see Table~1 for a summary).
\begin{figure*}
\epsscale{.9}
\plotone{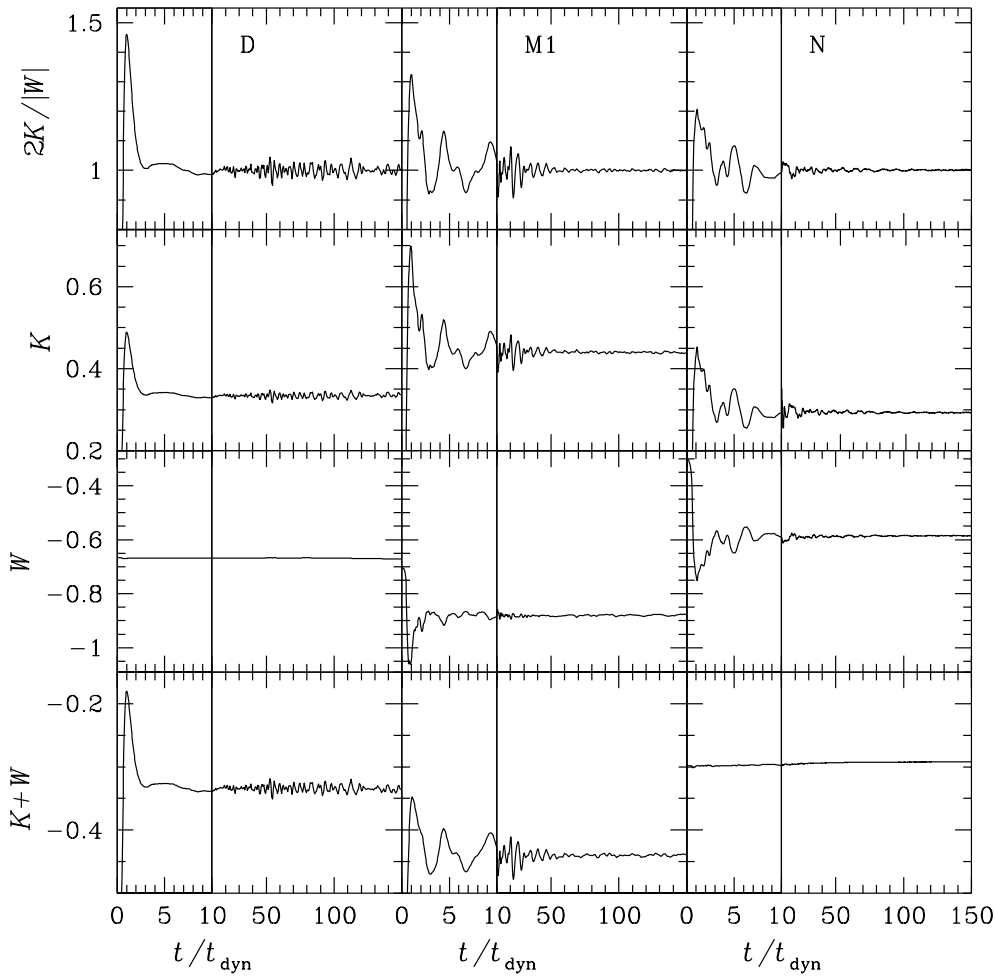}
\caption{Time evolution of $2 K/|W|$, $K$, $W$, and $K+W$ for
  simulations D, M1, and N.  $K$, $W$, and $K+W$ are in units of $\Estard$
  (left column), and $\Estarn$ (central and right columns). For clarity,
  the time axes are zoomed-in between 0 and 10.}
\label{figtime}
\end{figure*}

\subsection{Initial conditions and analysis of the simulations}
\label{secic}

We performed a set of five dissipationless-collapse N-body simulations,
starting from the same phase-space configuration: the initial particle
distribution follows the Plummer (1911) 
spherically symmetric density distribution
\begin{equation}
\label{eqplum}
\rho(r)={{3 \Mstar  \rstar^2 }\over 4 {\pi (r^2 +\rstar^2)^{5/2}}},
\end{equation}
where $\Mstar$ is the total mass and $\rstar$ a
characteristic radius.  
The choice of a Plummer sphere as initial condition is quite
artificial, and not necessarily the most realistic to reproduce
initial conditions in the cosmological context (e.g., Gunn \&
Gott~1972). We adopt such a distribution to adhere to other papers
dealing with collisionless collapse (e.g., Londrillo et al. 1991;
NLC06; see also Section~\ref{secdis}, in which we present the results
of a set of simulations starting from different initial conditions).
The particles are at rest, so the initial virial ratio
$2K/|W|=0$.  What is different in each simulation is the adopted
gravitational potential, which is Newtonian in simulation N, dMOND in
simulation D, and MOND with acceleration ratio $\kappa$ in simulations
M$\kappa$ ($\kappa$=1, 2, 4). For each simulation we define the
dynamical time $\tdyn$ as the time at which the virial ratio $2K/|W|$
reaches its maximum value. In particular, we find $\tdyn\sim2\tstard$
in simulation D, and $\tdyn\sim2\tstarn$ in simulations N, M1, M2 and
M4.  We note that $\tdyn\sim{G \Mstar^{5/2}(2|K+W|)^{-3/2}}$ in
simulation~N.

Following NLC06, the particles are spatially distributed according to
equation (\ref{eqplum}) and then randomly shifted in position (up to
$\rstar/5$ in modulus). This artificial, small-scale "noise" is
introduced to enhance the phase mixing at the beginning of the
collapse, because the numerical noise is small, and the velocity
dispersion is zero (see also Section 4.2).  As such, these
fluctuations are not intended to reproduce any physical clumpiness.

All the simulations (realized with $N=10^6$ particles, and a grid with
$\Nr=64$, $\Nth=16$ and $\Nph=32$) are evolved up $t=150\tdyn$. In all
cases the modulus of the center of mass position oscillates around
zero with r.m.s $\lsim 0.1\rstar$; similarly, the modulus of the total
angular momentum oscillates around zero\footnote{As an experiment we
also ran a simulation, with the same initial conditions and parameter
$\kappa$ as M1, in which the force was calculated from
equation~(\ref{eqcurl}) imposing $\Sv=0$. In this simulation the
linear and angular momentum are strongly not conserved: for instance,
the center of mass is already displaced by $\sim7\rstar$ after
$\sim30\tdyn$.}  with r.m.s. $\lsim0.02$, in units of
$\rstar\Mstar\vstarn$ (simulations M$\kappa$ and N) and of
$\rstar\Mstar\vstard$ (simulation D). $K+W$ in the Newtonian
simulation and $W$ in the dMOND simulation are conserved to within
$2\%$ and $0.6\%$, respectively.  The volume-limited energy balance
equation (\ref{eqbaltext}) is conserved with an accuracy of $1\%$ in
MOND simulations, independently of the adopted $V_0$.  To estimate
possible numerical effects, we reran one of the MOND collapse
simulations (M1) using $N=2\times 10^6$, $\Nr=80$, $\Nth=24$, and
$\Nph=48$: we found that the end-products of these two simulations do
not differ significantly, as far as the properties relevant to the
present work are concerned.

The intrinsic and projected properties of the collapse end-products
are determined as in NLC06. In particular, the position of the center
of the system is determined using the iterative technique described by
Power et al.~(2003).  Following Nipoti et al.~(2002), we measure the
axis ratios $c/a$ and $b/a$ of the inertia ellipsoid (where $a$, $b$
and $c$ are the major, intermediate and minor axis) of the final
density distributions, their angle-averaged profile and half-mass
radius $\rhalf$.  We fitted the final angle-averaged density profiles
with the $\gamma$-model (Dehnen~1993; Tremaine et al.~1994)
\begin{equation}
\label{eqgamma}
\rho (r)= {\rho_0  \rc^4 \over r^{\gamma} (\rc +r)^{4-\gamma}},
\end{equation}
where the inner slope $\gamma$ and the break radius $\rc$ are free
parameters, and the reference density $\rho_0$ is fixed by the total
mass $\Mstar$. The fitting radial range is $0.06\,\lsim\,
r/\rhalf\,\lsim\, 10$.  In order to estimate the importance of
projection effects, for each end-product we consider three orthogonal
projections along the principal axes of the inertia tensor, measuring
the ellipticity $\epsilon=1-\sbe/\sae$, the circularized projected
density profile and the circularized effective radius
{$\Re\equiv\sqrt{\sae\sbe}$} (where {$\sae$} and {$\sbe$} are the
major and minor semi-axis of the effective isodensity ellipse).  We
fit (over the radial range $0.1 \, \lsim \, R/\Re \, \lsim \, 10$) the
circularized projected density profiles of the end-products with the
$R^{1/m}$ Sersic~(1968) law:
\begin{equation}\label{eqser}
I(R)=\Ie \,
\exp\left\{-b(m)\left[ \left( \frac{R}{\Re} \right)^{1/m} -1 \right]\right\},
\end{equation}
where $\Ie\equiv I(\Re)$ and $b(m)\simeq 2m-1/3+4/405m$ (Ciotti \&
Bertin 1999).  In the fitting procedure $m$ is the only free
parameter, because $\Re$ and $\Ie$ are determined by their measured
values obtained by particle count.  In addition, we measure the
central velocity dispersion $\sgz$, obtained by averaging the
projected velocity dispersion over the circularized surface density
profile within an aperture of $\Re/8$. Some of these structural
parameters are reported in Table~2 for the five simulations described
above, as well as for three additional simulations, which start from
different initial conditions (see Section~\ref{secdis}).

\section{Results}
\label{secres}

In Newtonian gravity, collisionless systems reach virialization
through violent relaxation in few dynamical times, as predicted by the
theory (Lynden-Bell~1967) and confirmed by numerical simulations
(e.g. van Albada~1982). On the other hand, due to the non linearity of
the theory and the lack of numerical simulations, the details of
relaxation processes and virialization in MOND are much less
known. Thus, before discussing the specific properties of the collapse
end-products we present a general overview of the time evolution of
the virial quantities in our simulations, postponing to Section
\ref{secphsp} a more detailed description of the phase-space
evolution.  In particular, in Fig.~\ref{figtime} we show the time
evolution of $2 K/|W|$, $K$, $W$, and $K+W$ for simulations D, M1, and
N. In the diagrams time is normalized to $\tdyn$, so plots referring
to different simulations are directly comparable (the values of
$\tdyn$ in time units for the five simulations are given in
Section~\ref{secic}).  In simulation N (right column) we find the well
known behavior of Newtonian dissipationless collapses: $2 K/|W|$ has a
peak, then oscillates, and eventually converges to the equilibrium
value $2 K/|W|=1$; the total energy $K+W$ is nicely conserved during
the collapse, though it presents a secular drift, a well known feature
of time integration in N-body codes.  The time evolution of the same
quantities is significantly different in a dMOND simulation (left
column). In particular, the virial ratio $2 K/|W|$ quickly becomes
close to one, but is still oscillating at very late times because of
the oscillations of $K$, while $W$ is constant as expected.  As we
show in Section \ref{secphsp}, these oscillations are related to a
peculiar behavior of the system in phase space.  Finally, simulation
M1 (central column) represents an intermediate case between models N
and D: the system starts as dMOND, but soon its core becomes
concentrated enough to enter the Newtonian regime. After the initial
phases of the collapse, Newtonian gravity acts effectively in damping
the oscillations of the virial ratio. Overall, it is apparent how the
system is in a ``mixed'' state, neither Newtonian ($K+W$ is not
conserved) nor dMOND ($W$ is not constant).

\subsection{Properties of the collapse end-products}

\begin{table*}
 \flushleft{
  \caption{End-product properties.}
  \begin{tabular}{lrrrrrrrrrrr}
 & $\kappa$ & $c/a$ & $b/a$ &  $\gamma\;\;\;\;\;$ & $\rc/\rhalf\;\;\;\;$ & $m_a\;\;\;\;$ & $m_b\;\;\;\;$ & $m_c\;\;\;\;$ & $\epsilon_a$ &
$\epsilon_b$ &  $\epsilon_c$ \\
\hline
D  & - & 0.21 & 0.41 & $0.17^{+0.39}_{-0.17}$ & $0.27^{+0.09}_{-0.02}$  & $2.87\pm0.01$ & $2.50\pm0.03$ & $2.16\pm0.02$& 0.48 & 0.80 & 0.58\\
M1 & 1 & 0.47 & 0.85 &$1.24^{+0.40}_{-0.36}$ & $0.44^{+0.20}_{-0.12}$ & $3.20\pm0.07$ & $3.00\pm0.09$ & $3.07\pm0.13$& 0.42 & 0.51 & 0.17\\
M2 & 2 & 0.48 & 0.92 & $1.45^{+0.26}_{-0.34}$ & $0.53^{+0.19}_{-0.13}$& $3.38\pm0.08$ & $3.24\pm0.08$ & $3.28\pm0.12$& 0.49 & 0.45 & 0.08\\
M4 & 4 & 0.47 & 0.90 & $1.54^{+0.26}_{-0.32}$ & $0.58^{+0.20}_{-0.15}$ & $3.55\pm0.10$ & $3.40\pm0.11$ & $3.34\pm0.15$& 0.51 & 0.45 & 0.10\\
N  & -  & 0.45 & 0.91 &$1.69^{+0.13}_{-0.15}$& $0.74^{+0.14}_{-0.12}$ & $4.21\pm0.07$ & $4.35\pm0.08$ & $3.96\pm0.13$& 0.48 & 0.55 & 0.12\\
\hline
$\Dprime$  & - & 0.25 & 0.45 & $0.72^{+0.36}_{-0.50}$ & $0.37^{+0.12}_{-0.11}$  & $3.06\pm0.06$ & $2.90\pm0.04$ & $2.71\pm0.08$& 0.44 & 0.76 & 0.56\\
$\Mprime$ & 20 & 0.42 & 0.83 &$1.26^{+0.44}_{-0.40}$ & $0.47^{+0.25}_{-0.15}$ & $3.41\pm0.09$ & $3.36\pm0.06$ & $3.20\pm0.13$& 0.49 & 0.57 & 0.16\\
$\Nprime$  & -  & 0.45 & 0.93 &$1.78^{+0.15}_{-0.18}$& $0.78^{+0.24}_{-0.18}$ & $4.29\pm0.10$ & $4.56\pm0.15$ & $4.19\pm0.22$& 0.51 & 0.55 & 0.09\\
\hline
\end{tabular}
}
\medskip

\flushleft{First column: name of the simulation. $\kappa= G \Mstar
/\rstar^2\az$: acceleration ratio.  $c/a$ and $b/a$: minor-to-major
and intermediate-to-major axis ratios. $\gamma$, $\rc$: best-fit
$\gamma$-model parameters. $m_a$, $m_b$, $m_c$ and $\epsilon_a$,
$\epsilon_b$, $\epsilon_c$: best-fit Sersic indices and ellipticities
for projections along the principal axes.}

\end{table*}

\subsubsection{Spatial and projected density profiles}

\begin{figure*}
\epsscale{.9} 
\plotone{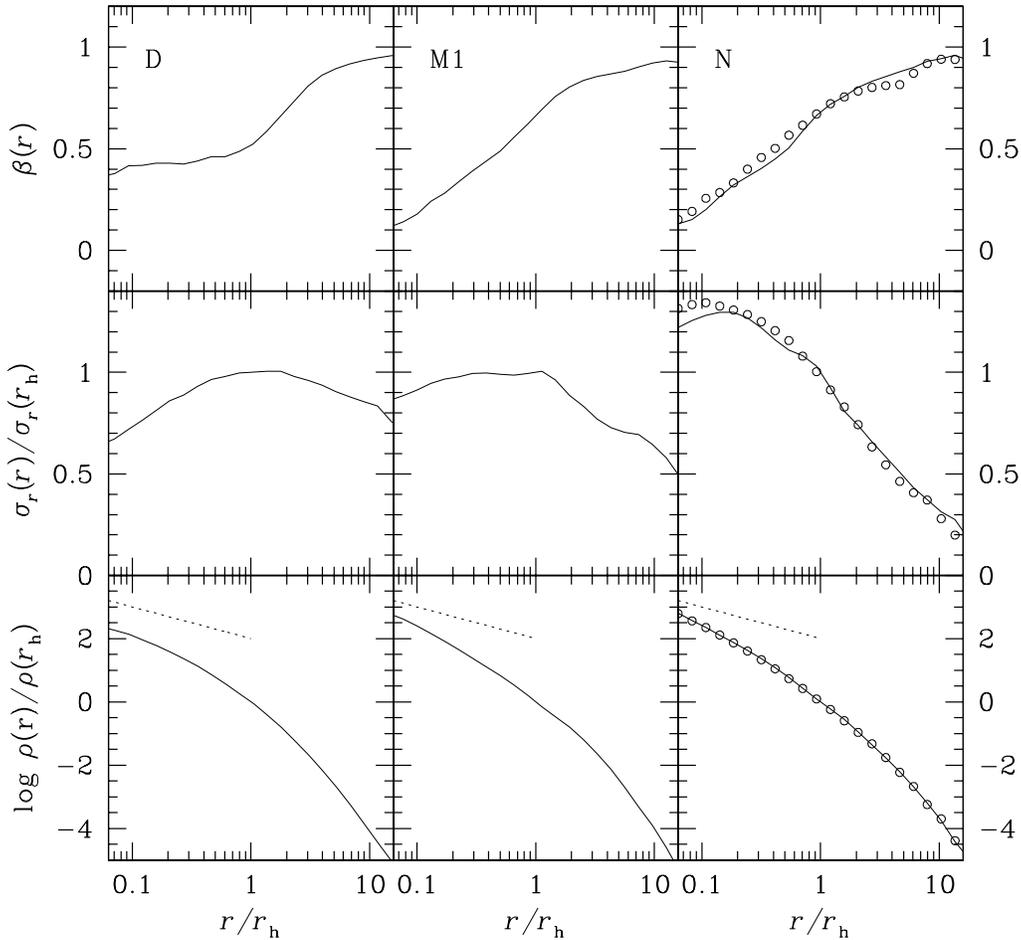}
\caption{Angle-averaged density, radial velocity-dispersion and
  anisotropy-parameter profiles (from bottom to top) for the
  end-products of simulations D, M1, and N. Dotted lines in the bottom
  panels represent $\rho \propto r^{-1}$ profiles, which are shown for
  reference. Empty circles in the right column show the corresponding
  profiles obtained with the FVFPS treecode from the same initial
  conditions.}
\label{fig3d}
\end{figure*}
\begin{figure*}
\epsscale{.9}
\plotone{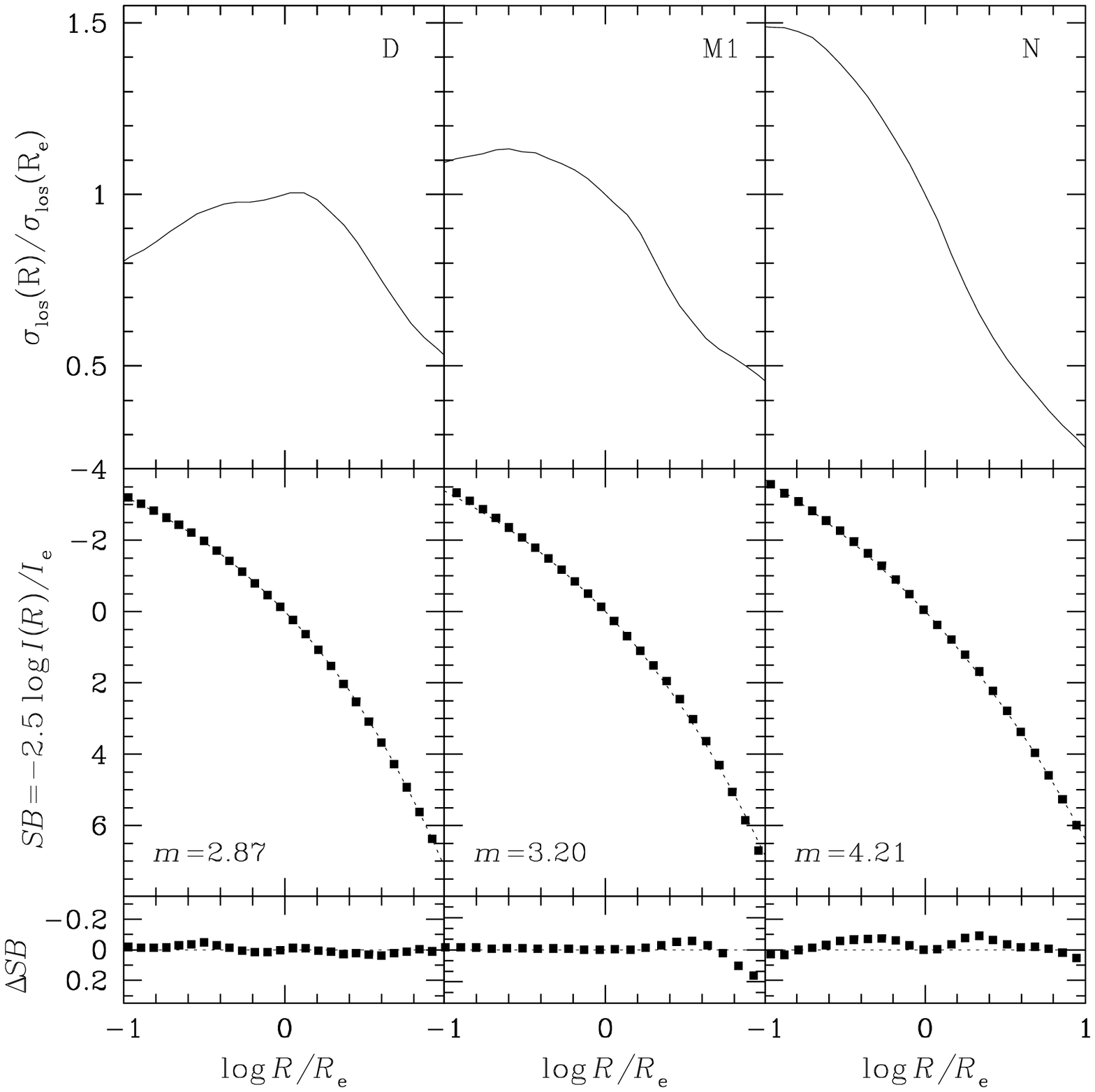}
\caption{ Line-of-sight velocity-dispersion profiles (top),
circularized projected density profiles and residuals of the Sersic
fit (bottom) of the end-products of simulations D, M1, and N (squares;
1-$\sigma$ error bars are always smaller than the symbol size).  The
dotted lines are the best-fitting Sersic models.}
\label{fig2d}
\end{figure*}

We found that all the simulated systems, once virialized, are not
spherically symmetric. However, while the dMOND collapse end-product
is triaxial ($c/a\sim0.2$, $b/a\sim0.4$), MOND and Newtonian
end-products are oblate ($c/a \sim c/b \sim 0.5$).  The ellipticity
$\epsilon$ of the projected density distributions (measured for each
of the principal projections) is found in the range $0.5-0.8$ in D,
and $0 - 0.5$ in M1, M2, M4 and N.  These values are consistent with
those observed in real ellipticals, with the exception of $\epsilon_b$
in model D (see Table 2), which would correspond - if taken at face
value - to an E8 galaxy. These result could be just due to the
procedure adopted to measure the ellipticity (see Section
\ref{secic}), however we find interesting that dMOND gravity could be
able to produce some system that would be unstable in Newtonian
gravity. We remark that a similar result, in the different context of
disk stability in MOND, has been obtained by Brada \& Milgrom (1999).

In order to describe the radial mass distribution of the final
virialized systems, we fitted their angle-averaged density profiles
with the $\gamma$-model (\ref{eqgamma}) over the radial range
$0.06\,\lsim\, r/\rhalf\,\lsim\, 10$. The best-fit $\gamma$ and $\rc$
for the final distribution of each simulation are reported in Table~2
together with their $1\sigma$ uncertainties (calculated from $\Delta
\chi^2=2.30$ contours in the space $\gamma-\rc$). As also apparent
from Fig.~\ref{fig3d} (bottom), the Newtonian collapse produced the
system with the steepest inner profile ($\gamma\sim1.7$), the dMOND
end-product has inner logarithmic slope close to zero, while MOND
collapses led to intermediate cases, with $\gamma$ ranging from $\sim
1.2$ ($\kappa=1$) to $\sim1.5$ ($\kappa=4$). We also note that the
ratio $\rc/\rhalf$ (indicating the position of the knee in the density
profile) increases systematically from dMOND to Newtonian simulations.

The circularized projected density profiles of the end-products are
analyzed as described in Section~\ref{secic}.  The best-fit Sersic
indices $m_a$, $m_b$ and $m_c$ (for projections along the axes $a$,
$b$, and $c$, respectively) are reported in Table~2, together with the
$1\sigma$ uncertainties corresponding to $\Delta \chi^2=1$; the
relative uncertainties on the best-fit Sersic indices are in all cases
smaller than 5 per cent and the average residuals between the data and
the fits are typically $0.05 \lsim\langle\Delta{SB}\rangle \lsim 0.2$,
where $SB\equiv-2.5 \log [I(R)/\Ie]$.  The fitting radial range
$0.1\,\lsim\, R/\Re\,\lsim\, 10$ is comparable with or larger than the
typical ranges spanned by observations (e.g., see Bertin, Ciotti \&
Del Principe~2002). In agreement with previous investigations, we
found that the Newtonian collapse produced a system well fitted by the
de Vaucouleurs~(1948) law. MOND collapses led to systems with Sersic
index $m<4$, down to $m\sim2$ in the case of the dMOND
collapse. Figure~\ref{fig2d} (bottom) shows the circularized
(major-axis) projected density profiles for the end-products of
simulations D, M1 and N together with their best-fit Sersic laws
($m=2.87$, $m=3.20$, and $m=4.21$, respectively), and the
corresponding residuals. Curiously, NLC06 found that low-$m$ systems
can be also obtained in Newtonian dissipationless collapses in the
presence of a pre-existing dark-matter halo, with Sersic index value
decreasing for increasing dark-to-luminous mass ratio.

\subsubsection{Kinematics}
\label{seckin}

We quantify the internal kinematics of the collapse end-products by
measuring the angle-averaged radial and tangential components
($\sigma_r$ and $\sigma_{\rm t}$) of their velocity-dispersion tensor,
and the anisotropy parameter $\beta(r) \equiv 1 -0.5\sigma^2_{\rm
t}/\sigma^2_r$. These quantities are shown in Fig.~\ref{fig3d} for
simulations D, M1, and N. We note that the $\sigma_r$ profile
decreases more steeply in the Newtonian than in the MOND end-products,
while it presents a hole in the inner regions of the dMOND system. In
addition, the dMOND galaxy is radially anisotropic ($\beta\sim0.4$)
even in the central regions, where models N and M1 are approximately
isotropic ($\beta \sim 0.1$).  All systems are strongly radially
anisotropic for $r\gsim\rhalf$.  For each model projection we computed
the line-of-sight velocity dispersion $\sglos$, considering particles
in a strip of width $\Re/4$ centered on the semi-major axis of the
isophotal ellipse. The line-of-sight velocity-dispersion profiles (for
the major-axis projection) are plotted in the top panels of
Fig.~\ref{fig2d}. The Newtonian profile is very steep within $\Re$,
while MOND and dMOND profiles are significantly flatter there.  As
well as $\sigma_r$, $\sglos$ decreases for decreasing radius in the
inner region of model D.  The kinematical properties of M2 and M4 are
intermediate between those of M1 and of N: overall we find only weak
dynamical non-homology among MOND end-products. The empty symbols in
Fig. \ref{fig3d} (right column) refer to a test Newtonian simulation
run with the FVFPS treecode (with $4\times10^5$ particles). The
structural and kinematical properties of the end-product of this
simulation are clearly in good agreement with those of the end-product
of simulation N (solid lines), which started from the same initial
conditions.

\subsection{Phase-space properties of MOND collapses}
\label{secphsp}

\begin{figure*}
\epsscale{.9} 
\plotone{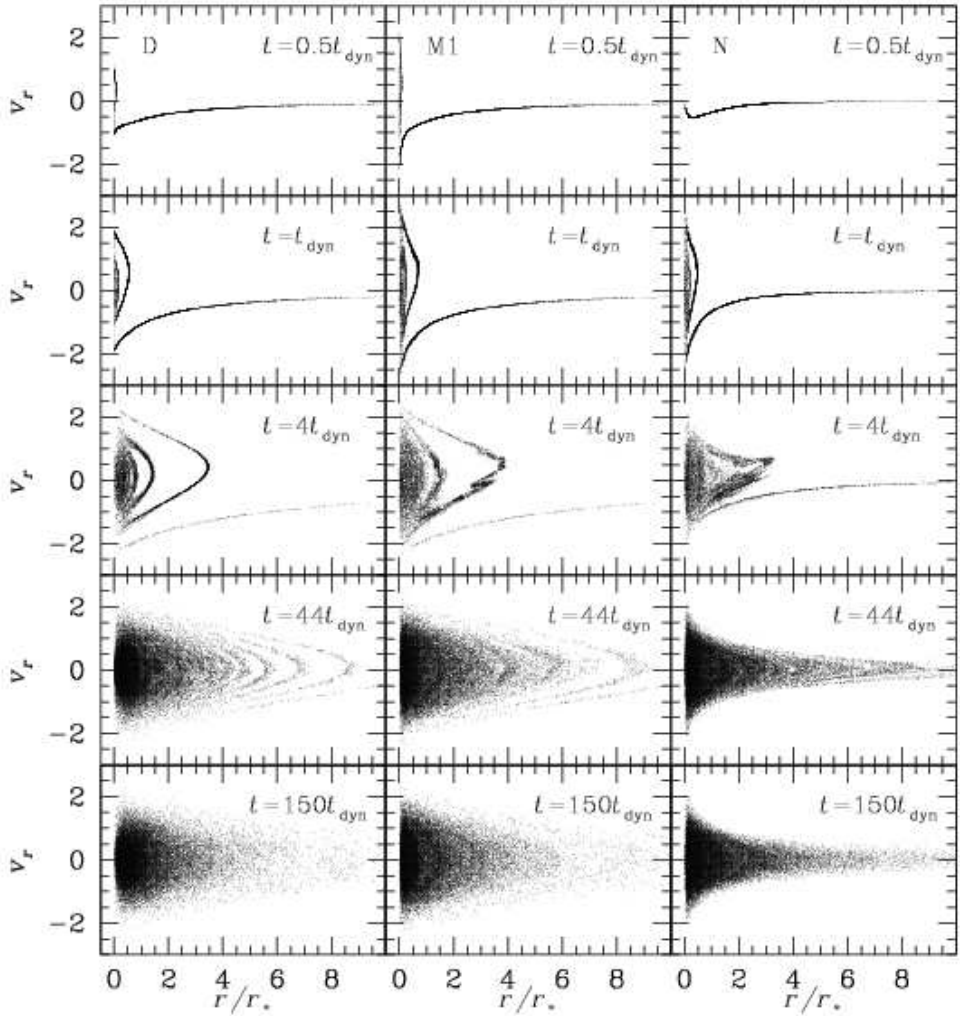}
\caption{Phase-space (radial-velocity vs. radius) diagrams for
  simulations D, M1, and N at various times. $\vr$ is in units of
  $\vstard$ (simulation D), and $\vstarn$ (simulations M1 and N; see
  Table~1).}
\label{figphsp}
\end{figure*}

\begin{figure*}
\epsscale{.9}
\plotone{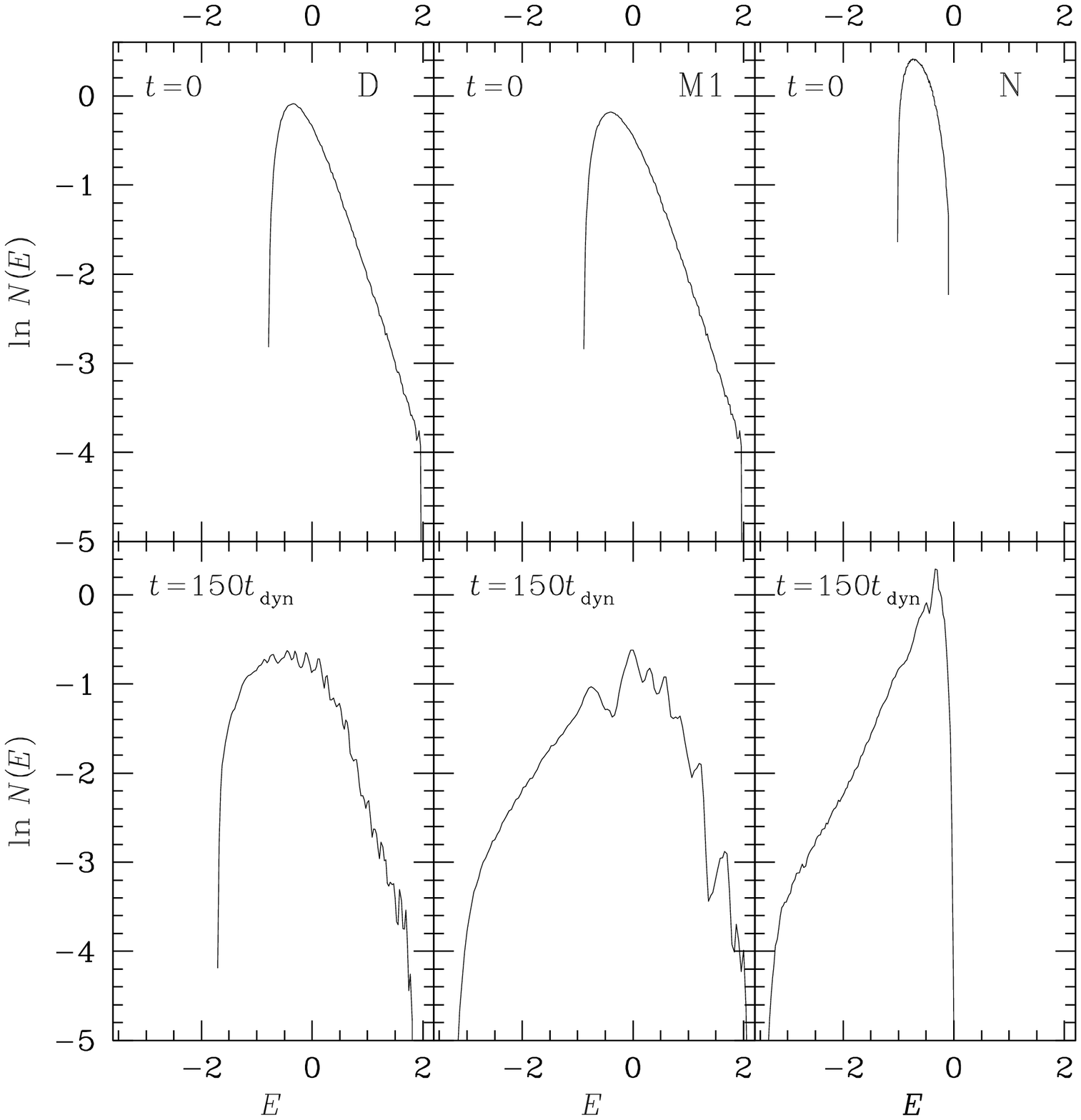}
\caption{Initial (top) and final (bottom) differential energy
distributions. The energy per unit mass $E$ is in units of
$\Estard/\Mstar$ (model D), and $\Estarn/\Mstar$ (models M1 and
N). The energy zero points in models D and M1 are such that the most
bound particles of the M1 and N end-products have the same energy, and
the highest energy particles of models D and M1 at $t=0$ have the same
energy.}
\label{fignde}
\end{figure*}

To explore the phase-space evolution of the systems during the
collapse and the following relaxation we consider time snapshots of
the particles radial velocity ($\vr$) vs. radius as in Londrillo et
al.~(1991). In Fig.~\ref{figphsp} we plot five of these diagrams for
simulations D, M1 and N: each plot shows the phase-space coordinates
of 32000 particles randomly extracted from the corresponding
simulation, and, as in Fig.~\ref{figtime}, times are normalized to the
dynamical time $\tdyn$ (see Section~\ref{secic}).  At time
$t=0.5\tdyn$ all particles are still collapsing in simulation N, while
in MOND simulations a minority of particles have already crossed the
center of mass, as revealed by the vertical distribution of points at
$r\sim0$ in panels D and M1. At $t=\tdyn$ (time of the peak of $2
K/|W|$ in the three models), sharp shells in phase space are present,
indicating that particles are moving in and out collectively and phase
mixing has not taken place yet.  At $t=4\tdyn$ is already apparent
that phase mixing is operating more efficiently in simulation N than
in simulation M1, while there is very little phase mixing in the dMOND
collapse.  At significantly late times ($t=44\tdyn$), when the three
systems are almost virialized ($2 K/|W|\sim 1$; see
Fig.~\ref{figtime}), phase mixing is complete in simulation N, but
phase-space shells still survive in models M1 and D.  Finally, the
bottom panels show the phase-space diagrams at equilibrium
($t=150\tdyn$), when phase mixing is completed also in the MOND and
dMOND galaxies: note that the populated region in the ($r$,$\vr$)
space is significantly different in MOND and in Newtonian gravity,
consistently with the sharper decline of radial velocity dispersion in
the Newtonian system. 

Thus, our results indicate that phase mixing is
more effective in Newtonian gravity than in MOND\footnote{Ciotti,
Nipoti \& Londrillo (2007) found similar results in ``ad hoc''
numerical simulations in which the angular force components were
frozen to zero, so that the evolution was driven by radial forces only. 
In fact, while phase mixing is less effective both in MOND 
and in Newtonian simulations with respect to the simulations here reported, 
the phase mixing 
time scale in MOND is still considerably longer than in Newtonian gravity.}. 
It is
then interesting to estimate in physical units the phase-mixing
timescales of MOND systems. From Table~1 it follows that
$\tstarn\simeq4.7\,(\rstar/\kpc)^{3/2}(\Mstar/10^{10}
\Msun)^{-1/2}\Myr=29.8\kappa^{-3/4}(\Mstar/10^{10} \Msun)^{1/4}\Myr$
for $\az=1.2\times 10^{-10} {\rm m}\,{\rm s}^{-2}$. For example, in
the case of model M1, adopting $\Mstar=10^{12} \Msun$ (and
$\rstar=\sqrt{G\Mstar/\az}\simeq34\kpc$), shells in phase space are
still apparent after $\sim8.3\Gyr$ ($\simeq44\tdyn$).  Simulation M1
might also be interpreted as representing a dwarf elliptical galaxy
of, say, $\Mstar= 10^9 \Msun$ (and $\rstar=\sqrt{G\Mstar/\az}\simeq
1.1\kpc$). In this case $44\tdyn\sim 1.5 \Gyr$. We conclude that in
some MOND systems substructures in phase space can survive for
significantly long times.

In addition to the ($r$,$\vr$) diagram, another useful diagnostic to
investigate phase-space properties of gravitational systems is the
energy distribution $N(E)$ (i.e. the number of particles with energy
per unit mass between $E$ and $E+dE$; e.g., Binney \& Tremaine~1987;
Trenti \& Bertin~2005).  Independently of the force law, the energy
per unit mass of a particle orbiting at $\xv$ with speed $v$ in a
gravitational potential $\phi(\xv)$ is $E=v^2/2+\phi(\xv)$, and $E$ is
constant if $\phi$ is time-independent.  In Newtonian gravity $\phi$
is usually set to zero at infinity for finite-mass systems, so $E<0$
for bound particles; in MOND all particles are bound, independently of
their velocity, because $\phi$ is confining, and all energies are
admissible. This difference is reflected in Fig.~\ref{fignde}, which
plots the initial (top) and final (bottom) differential energy
distributions for simulations D, M1, and N.  Given that the particles
are at rest at $t=0$, the initial $N(E)$ depends only on the structure
of the gravitational potential, and is significantly different in the
Newtonian and MOND cases. We also note that $N(E)$ is basically the
same in models D and M1 at $t=0$, because model M1 is initially in
dMOND regime. In accordance with previous studies, in the Newtonian
case the final differential $N(E)$ is well represented by an
exponential function over most of the populated energy range
(Binney~1982; van Albada~1982; Ciotti~1991; Londrillo et al.~1991;
NLC06).  In contrast, in model D the final $N(E)$ decreases for
increasing energy, qualitatively preserving its initial shape. In the
case of simulation M1 it is apparent a dichotomy between a Newtonian
part at lower energies (more bound particles), where $N(E)$ is
exponential, and a dMOND part at higher energies, where the final
$N(E)$ resembles the initial one. We interpret this result as another
manifestation of a less effective phase-space reorganization in MOND
than in Newtonian collapses.

\subsection{Comparison with the observed scaling relations of elliptical galaxies}
\label{secsca}

\begin{figure*}
\epsscale{.9}  
\plotone{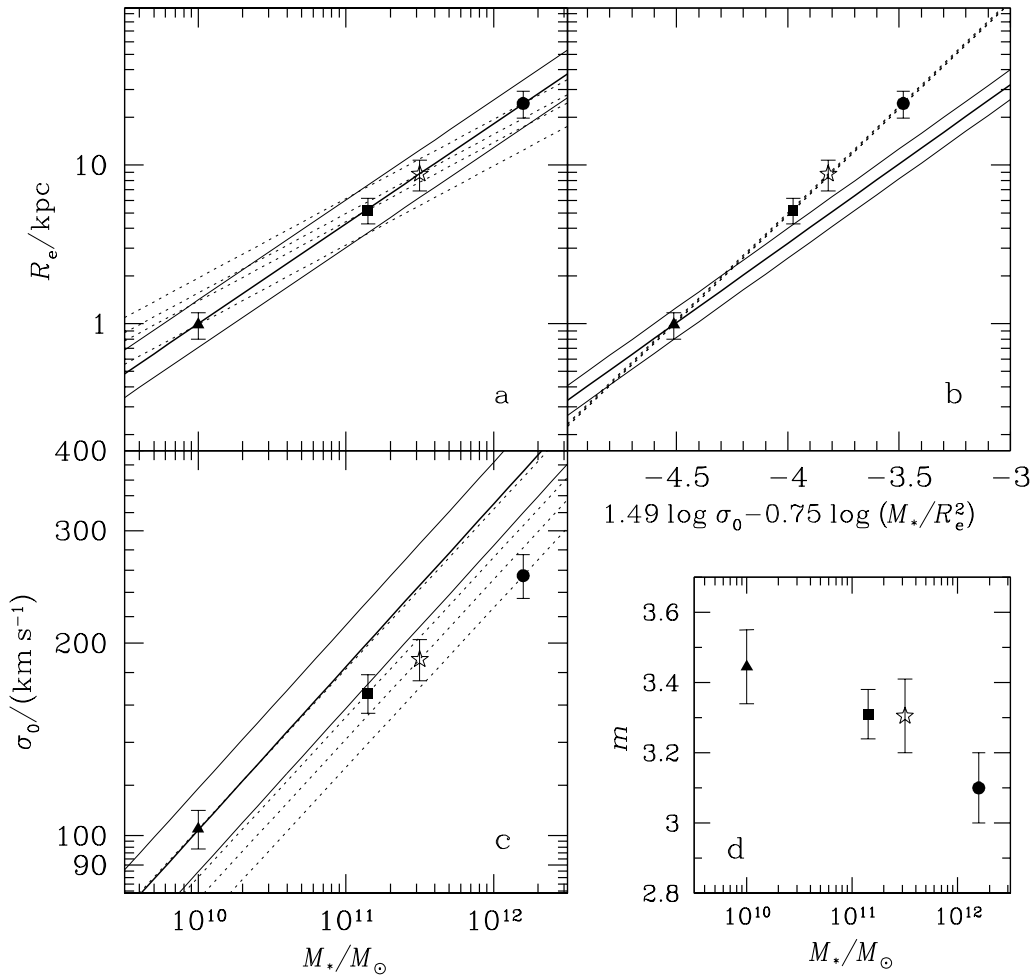}
\caption{The location of the end-products M1 (circles), M2 (squares),
  M4 (triangles) and $\Mprime$ (stars) in the planes $\Mstar-\Re$ (a),
  $\Mstar-\sigma_0$ (c), $\Mstar-m$ (d), and in the plane in which the
  FP is seen edge-on (b). Vertical bars account for the projection
  effects, while the solid symbols refer to average values between the
  two extremes.  The thick solid lines represent the observed Kormendy
  (a), FJ (c), and FP (b) relations, and the thin solid lines give
  estimates of the corresponding scatter (Bernardi et al.~2003ab).
  The normalization is such that $\Re\simeq 4 \kpc$ for stellar mass
  $\Mstar=10^{11}\Msun$ (Shen et al.~2003). See text for the meaning
  of the dotted lines.}
\label{figfp}
\end{figure*}

It is not surprising that galaxy scaling relations represent an even
stronger test for MOND than for Newtonian gravity, due to the absence
of dark matter and the existence of the critical acceleration $\az$
with a universal value in the former theory (e.g., see Milgrom~1984;
Sanders~2000). For example, when interpreting the FP tilt in Newtonian
gravity one can invoke a systematic and fine-tuned increase of the
galaxy dark-to-luminous mass ratio with luminosity (e.g., Bender,
Burstein \& Faber~1992; Renzini \& Ciotti~1993; Ciotti, Lanzoni \&
Renzini~1996), while in MOND the tilt should be related to the
characteristic acceleration $\az$. Note, however, that in MOND as well
as in Newtonian gravity other important physical properties may help
to explain the FP tilt, such as a systematic increase of radial
orbital anisotropy with mass or a systematic structural weak homology
(Bertin et al.~2002). Due to the relevance of the subject, we attempt
here to derive some preliminary hints.  In particular, for the first
time, we can compare with the scaling relations of elliptical galaxies
MOND models produced by a formation mechanism, yet as simple as the
dissipationless collapse.

In this Section we consider the end-products of simulations M1, M2,
and M4.  As already discussed in Section~\ref{secsim}, each of the
three systems corresponds to a family with constant
$\Mstar/\rstar^2$. This degeneracy is represented by the straight
dotted lines in Fig.~\ref{figfp}a: all galaxies on the same dotted
line have the same $\kappa$ value.  This behavior is very different
from the Newtonian case, in which the result of a N-body simulation
can be placed anywhere in the space $\Re-\Mstar$, by arbitrarily
choosing $\Mstar$ and $\rstar$.  For comparison with observations, the
specific scaling laws represented in Fig.~\ref{figfp} (thick solid
lines) are the near-infrared $z^*$-band Kormendy relation $\Re \propto
\Mstar^{0.63}$ and FJ relation $\Mstar\propto\sigma_0^{3.92}$
(Bernardi et al.~2003a), and the edge-on FP relation in the same band
$\log \Re = A \log \sigma_0 + B \log (\Mstar / \Re^2) +const$ (with
$A=1.49$, $B=- 0.75$; Bernardi et al. 2003b), under the assumption of
luminosity-independent mass-to-light ratio.

The physical properties of each model are determined as
follows. First, for each model (identified by a value of $\kappa= G
\Mstar/\rstar^2\az$) we measure the ratio $\Re/\rstar$ (see
Section~\ref{secic}), and so we obtain $\Re=\Re(\kappa,\Mstar)$. This
function, for fixed $\kappa$ and variable $\Mstar$, is a dotted line
in Fig.~\ref{figfp}a. As apparent, only one pair $(\Re,\Mstar)$
satisfies the Kormendy relation for each $\kappa$: in particular, we
obtain that models M1, M2, and M4 have stellar masses $1.6\times
10^{12}$, $1.4\times 10^{11}$, and $10^{10}\Msun$, respectively (so
lower $\kappa$ models correspond to higher mass systems).  We are now
in the position to obtain $\rstar=\sqrt{G\Mstar/\az\kappa}$ and
$\vstarn=(\kappa\az G\Mstar)^{1/4}$, so we know the physical value of
the projected central velocity dispersion, and we can place our models
also in the FJ and FP planes.  It is apparent that these two relations
are not reproduced, in particular by massive galaxies.  We note that
this discrepancy cannot be fixed even when considering the mass
interval allowed by the scatter in the Kormendy relation (thin solid
lines in panel a), as revealed by the dotted lines in
Fig.~\ref{figfp}bc, which are just the projections of the dotted lines
in Fig.~\ref{figfp}a onto the planes of the edge-on FP\footnote{In
Fig.~\ref{figfp}b the dotted lines are nearly coincident because
1) the  models are almost homologous, and 2) the variable in
abscissa is independent of $\kappa$, being the FP coefficients
$A\sim-2B$ in this case.} and the FJ. Finally, Fig.~\ref{figfp}d plots
the best-fit Sersic index $m$ of the models as a function
of $\Mstar$.  Observations show that elliptical galaxies are
characterized by $m$ values increasing with size: $m\gsim 4$ for
galaxies with $\Re\gsim3\kpc$ and $m\lsim 4$ for those with
$\Re\lsim3\kpc$ (e.g. Caon, Capaccioli \& d'Onofrio~1993). Our models
behave in the opposite way, as $m$ decreases for increasing size
(mass) of the system. So, while model M4 ($\Re\sim 1\kpc$, $m\sim3.4$)
is consistent with observations, models M1 and M2 have significantly
lower $m$ than real ellipticals of comparable size. However, this
finding is not a peculiarity of MOND gravity: also in Newtonian
gravity dissipationless collapse end-products with $m>4$ are obtained
only for specific initial conditions (NLC06), while equal mass
Newtonian mergings are able to produce high-$m$ systems (Nipoti et
al.~2003).

So far we have compared the results of our simulations with the
scaling relations of high-surface brightness galaxies. However, it is
well known that low surface-brightness hot stellar systems, such as
dwarf ellipticals and dwarf spheroidals, have larger effective radii
than predicted by the Kormendy relation (e.g., Bender et al.~1992;
Capaccioli, Caon \& d'Onofrio~1992; Graham \& Guzm\'an~2003). In
particular, dwarf ellipticals are characterized by effective surface
densities comparable to those of the most luminous ellipticals, while
their surface brightness profiles are characterized by Sersic indices
smaller than $4$ (e.g. Caon et al.~1993; Trujillo, Graham, \&
Caon~2001).  Dwarf spheroidals are the lowest surface-density stellar
systems known, and typically have exponential ($m\sim1$) luminosity
profiles (e.g. Mateo~1998).  So, simulations M1 and D can be
interpreted as modeling a dwarf elliptical and a dwarf spheroidal,
respectively, and their end-products qualitatively reproduce the
surface brightness profiles of the observed systems. As pointed out in
Section~\ref{seckin}, the velocity-dispersion profile of model D is
rather flat, with a hole in the central regions: interestingly
observations of dwarf spheroidals indicate that their
velocity-dispersion profiles are also flat (e.g. Walker et al.~2006
and references therein).

\section{Discussion and conclusions}
\label{secdis}

In this paper we studied the dissipationless collapse in MOND by using
a new three-dimensional particle-mesh N-body code, which solves the
MOND field equation~(\ref{eqMOND}) exactly.  For obvious computational
reasons, we did not attempt a complete exploration of the parameter
space, and we just presented results of a small set of numerical
simulations, ranging from Newtonian to dMOND systems. The main results
of the present study can be summarized as follows:
\begin{itemize}
\item The intrinsic structural and kinematical properties of the MOND
  collapse end-products depend weakly on their characteristic surface
  density: lower surface-density systems have shallower inner density
  profile, flatter velocity-dispersion profile, and more radially
  anisotropic orbital distribution than higher surface-density
  systems.
\item The projected density profiles of the MOND collapse end-products
  are characterized by Sersic index $m$ lower than $4$, and decreasing
  for decreasing mean surface density.  In particular, the end-product
  of the dMOND collapse, modeling a very low surface density system,
  is characterized by a Sersic index $m\sim 2$ and by a central hole
  in the projected velocity-dispersion profile.
\item We found impossible to satisfy simultaneously the observed
  Kormendy, Faber-Jackson and Fundamental Plane relations of
  elliptical galaxies with the MOND collapse end-products, under the
  assumption of a luminosity independent mass-to-light ratio. In other
  words, this point and the two points above show that, in the
  framework of dissipationless collapse, the presence of a
  characteristic acceleration is not sufficient to reproduce important
  observed properties of spheroids of different mass and surface
  density, such as their scaling relations and weak structural
  homology.
\item From a dynamical point of view we found that phase mixing is
  less effective (and stellar systems take longer to relax) in MOND
  than in Newtonian gravity.
\end{itemize}

A natural question to ask is how the end-products of our simulations
would be interpreted in the context of Newtonian gravity. Clearly,
models D and N would represent dark-matter dominated and baryon
dominated stellar systems, respectively. More interestingly, models
M1, M2, and M4, once at equilibrium, would be characterized by a {\it
dividing radius}, separating a baryon-dominated inner region (with
accelerations higher than $\az$) from a dark-matter dominated outer
region (with accelerations lower than $\az$).  This radius is $\sim
1.1 \rhalf$ for M1, $\sim1.8 \rhalf$ for M2, and $\sim 2.7 \rhalf$ for
M4.  So, all these models would show little or no dark matter in their
central regions.  Remarkably, observational data indicate that there
is at most as much dark matter as baryonic matter within the effective
radius of ellipticals (e.g. Bertin et al.~1994; Cappellari et al.~2006
and references therein).

The conclusions above have been drawn by considering only simulations
starting from an inhomogeneous Plummer density distribution.  To
explore the dependence of these results on this specific choice, we
ran also three simulations starting from a cold ($2K/|W|=0$),
inhomogeneous and truncated density distribution
$\rho(r)=C\Mstar/(\rstar^3 +r^3)$, where $C^{-1}\equiv
4\pi\ln(1+\rt^3/\rstar^3)/3$, $\Mstar$ is the total mass, and
$\rt=20\rstar$ is the truncation radius. Inhomogeneities are
introduced as described in Section~\ref{secic}. Note that in the
external parts the new initial conditions are significantly flatter
than a Plummer sphere.  The three simulations are labeled $\Dprime$
(dMOND), $\Mprime$ (MOND with acceleration ratio\footnote{This value
of $\kappa$ is not directly comparable with those in simulations M1,
M2 and M4, because of the different role of $\rstar$ in the
corresponding initial distributions.}  $\kappa=20$) and $\Nprime$
(Newtonian).  As in the case of Plummer initial conditions, also in
these cases the final intrinsic and projected density distributions
are well represented by $\gamma$-models and Sersic models,
respectively.  In analogy with model N, the Newtonian collapse
$\Nprime$ produced the system with the steepest central density
distribution (see Table~2).  In addition, model $\Mprime$, when
compared with the scaling laws of ellipticals (stars in
Fig.~\ref{figfp}), follows the same trend as models M1, M2 and M4. The
analysis of the time-evolution in phase-space of models $\Dprime$,
$\Mprime$ and $\Nprime$ confirmed that mixing and relaxation processes
are less effective in MOND than in Newtonian gravity.

How do the presented results depend on the specific choice
(equation~\ref{eqmu}) of the MOND interpolating function $\mu$?
Recently, a few other interpolating functions have been proposed to
better fit galactic rotation curves (Famaey \& Binney~2005), and in
the context of TeVeS (Bekenstein~2004; Zaho \& Famaey~2006).  It is
reasonable to expect that the exact form of $\mu$ is not critical in a
violent dynamical process such as dissipationless collapse. We
verified that this is actually the case, by running an additional MOND
simulation with the same initial conditions and parameter $\kappa$ as
simulation M1, but adopting $\mu(y)=y/(1+y)$, as proposed by Famaey \&
Binney~(2005). In fact, neither in the time-evolution nor in the
structural and kinematical properties of the end-products we found
significant differences between the two simulations. This result
suggests that, in the context of structure formation in MOND, the
crucial feature is the presence of a characteristic acceleration
separating the two gravity regimes, while the details of the
transition region are unimportant.

Though the dissipationless collapse is a very simplistic model for
galaxy formation, it is expected to describe reasonably well the last
phase of ``monolithic-like'' galaxy formation, in which star formation
is almost completed during the initial phases of the collapse.  The
importance of gas dissipation in the formation of elliptical galaxies
is very well known, going back to the seminal works of Rees \&
Ostriker (1977) and White \& Rees (1978; see also Ciotti, Lanzoni \&
Volonteri 2006, and references therein, for a discussion of the
expected impact of gas dissipation on the scaling laws followed by
elliptical galaxies).  This aspect has been completely neglected in
our exploration, and we are working on an hybrid (stars plus gas)
version of the MOND code to explore quantitatively this issue.  We
also stress that the dissipationless collapse process catches the
essence of violent relaxation, which is certainly relevant to the
formation of spheroids even in more complicated scenarios, such as
merging. For example, it is well known that in Newtonian dynamics
systems with de Vaucouleurs profiles are produced by dissipationless
merging of spheroids (e.g. White~1978) or disk galaxies
(e.g. Barnes~1992) as well as by dissipationless collapses (van
Albada~1982).  Merging simulations in MOND have not been performed so
far, and a relevant and still open question is how efficient merging
is in MOND, in which the important effect of dark matter halos is
missing, and galaxies are expected to collide at higher speed than in
Newtonian gravity (Binney~2004; Sellwood~2004). Our results,
indicating that relaxation takes longer in MOND than in Newtonian
gravity, go in the direction of making merging time scales even longer
in MOND; on the other hand, analytical estimates seem to indicate
shorter dynamical friction time-scales in MOND than in Newtonian
gravity (Ciotti \& Binney~2004). So, the next application of our code
will be the study of galaxy merging in MOND.

\acknowledgments We are grateful to James Binney and Scott Tremaine
for helpful discussions and to the anonymous referee for useful
comments.  This work was partially supported by the MIUR grant
CoFin2004.

\appendix

\section{The volume-limited energy-balance equation in MOND}
\label{appetot}

In this Appendix we derive a useful volume-limited integral relation
representing energy conservation in MOND, well suited to test
numerical simulations.  The total (ordered and random) kinetic energy
per unit volume of a continuous distribution with density $\rho$ and
velocity field $\uvec$ is
\begin{equation}
k={\rho\over 2}(||\uvec||^2+\Tr{\sigma^2_{ij}}),
\end{equation}
where $\sigma^2_{ij}$ is the velocity-dispersion tensor. In the present
case the energy balance equation is (e.g. Ciotti~2000)
\begin{equation}
{d \over d t}\int_{V(t)}k\dxcube=-\int_{V(t)}\rho \uvec\cdot\nabla\phi\dxcube,
\label{eqbalone}
\end{equation}
where the integral in the r.h.s. is the work per unit time done by
mechanical forces.  By application of the Reynolds transport theorem
and using the mass continuity equation we obtain
\begin{equation}
{\partial \over \partial t}(k+\rho\phi)+\diverg
[(k+\rho\phi)\uvec]=\rho{\partial \phi \over \partial t}.
\label{eqbala}
\end{equation}
When $\phi$ is the MOND gravitational potential,  $\rho$ can be
eliminated using equation~(\ref{eqMOND}), so   
\begin{equation}
4 \pi G \rho{\partial \phi \over \partial t}=\diverg(\mu
  \nabla\phi){\partial \phi \over \partial t}=\diverg\left(\mu
  \nabla\phi {\partial \phi \over \partial t}\right)-{\az^2\over
  2}{\partial \over \partial t}
  \left[\mathcal{F}\left({||\nabla\phi||\over \az}\right)\right],
\end{equation}
where $\mathcal{F}$ is defined in equation~(\ref{eqeffe}).  Thus,
equation~(\ref{eqbala}) can be written as
\begin{equation}
{\partial \over \partial t}\left[k+\rho\phi+{\az^2\over 8 \pi
G}\mathcal{F}\left({||\nabla\phi||\over
    \az}\right)\right]+\diverg\left[(k+\rho\phi)\uvec-{\mu\nabla\phi\over 4\pi G}{\partial \phi \over \partial t}\right]=0.
\label{eqbalafin}
\end{equation}
By integration over a fixed control volume $V_0$ enclosing all the
system mass one obtains equation~(\ref{eqbaltext}).

\section{The virial trace $W$ in deep-MOND systems of finite mass}
\label{appw}

Here we prove that $W=-(2/3)\sqrt{G\az M_*^3}$ for any dMOND system of
finite mass $\Mstar$. Eliminating $\rho$ from equation (\ref{eqwij})
by using equation (\ref{eqdMOND}), and considering the trace of the
resulting expression one finds
\begin{equation}
W=-{1 \over 4\pi G \az}\int \Dcal[\phi] \nabla\cdot\left({\Vert\nabla\phi\Vert}\nabla\phi\right)\dxcube,
\label{eqwdiv}
\end{equation}
where we define the operator $\Dcal\equiv<\xv,\nabla>$.  The
remarkable fact behind the proof is that the integrand above can be
written as the divergence of a vector field, so only contributions
from $r \to \infty$ are important. We will then use the spherically
symmetric asymptotic behavior of dMOND solutions for $r \to \infty$
(BM)
\begin{equation}
\gv=-\nabla \phi \sim-{\sqrt{G \Mstar \az}\over r} \hat{\bf e}_r
\label{eqasym}
\end{equation}
and Gauss theorem to evaluate $W$.  

{\bf Theorem}. For a generic potential the following identity holds:
\begin{equation}
\Dcal[\phi]\nabla \cdot({\Vert\nabla\phi\Vert}\nabla\phi)=\nabla
\cdot \left(  \Dcal[\phi]{\Vert\nabla\phi\Vert}\nabla \phi -
      {\xv {\Vert\nabla\phi\Vert}^3  \over 3}\right).
\label{eqtheo}
\end{equation} 

{\it Proof}. From standard vector analysis (e.g. Jackson~1999) it follows that 
\begin{equation}
\Dcal[\phi]
\nabla\cdot\left({\Vert\nabla\phi\Vert}\nabla\phi\right)= 
\nabla\cdot\left(\Dcal[\phi]{\Vert\nabla\phi\Vert}\nabla\phi\right)-
{\Vert\nabla\phi\Vert}<\nabla\phi,\nabla\Dcal[\phi]>,
\label{eqxnabla}
\end{equation}
and
\begin{equation}
{\Vert\nabla\phi\Vert}<\nabla\phi,\nabla\Dcal[\phi]>={\diverg\left(
  \xv{\Vert\nabla\phi\Vert}^3\right)\over 3}.
\label{eqdivthird}
\end{equation}
Identity (\ref{eqdivthird}) follows from the expansion 
$\nabla \Dcal[\phi]=\nabla \phi+\Dcal[\nabla \phi]$ as
\begin{equation}
{\Vert\nabla\phi\Vert}^3+{\Vert\nabla\phi\Vert}<\nabla\phi,\Dcal[\nabla\phi]>={\Vert\nabla\phi\Vert}^3+{
  \Dcal\left[{\Vert\nabla\phi\Vert}^3\right]\over 3}={\diverg\left(\xv
  {\Vert\nabla\phi\Vert}^3\right)\over 3}.
\end{equation}
Combining equations (\ref{eqxnabla}) and (\ref{eqdivthird}) completes
the proof of equation~(\ref{eqtheo}).

We now transform the volume integral (\ref{eqwdiv}) in a surface
integral over a sphere of radius $r$, and we consider the limit for $r
\to \infty$ together with the asymptotic relation (\ref{eqasym}),
obtaining
\begin{equation}
W=-{1 \over 4\pi G \az} \lim_{r\to\infty} \int_{4\pi} {2 \over 3} r^3 g^2 d\Omega
=-{2 \over 3}\sqrt{G\az M_*^3}.
\end{equation}

\section{Scaling of the equations}
\label{appscal}

Given a generic density distribution $\rho$, and the mass and length
units $\Mstar$ and $\rstar$, we define the dimensionless quantities
$\txv\equiv \xv/\rstar$, and $\trho\equiv\rho\rstar^3/\Mstar$.  From
equation~(\ref{eqgv}) the equation of motion for a test particle can
be written in dimensionless form as
\begin{equation}
\label{eqmotion}
{d^2 \txv \over d \tilde{t}^2}=-{\phistar \tstar^2 \over \rstar^2}\tnabla \tphi,
\end{equation}
 where $\phistar$ and $\tstar$ are for the moment two unspecified
scaling constants, $\tphi=\phi/\phistar$ and $\tilde{t}=t/\tstar$, and
the dimensionless gradient operator is $\tnabla = \rstar \nabla $. In
all of our simulations we define $\tstar\equiv\rstar/\sqrt{\phistar}$,
so that the scaling factor in equation~(\ref{eqmotion}) is unity,
while $\phistar$ is specified case-by-case from the field equation as
follows.

In Newtonian gravity the Poisson equation~(\ref{eqPoisson}) can be
written as
\begin{equation}
\tnabla^2\tphi=4 \pi{G \Mstar \over \rstar \phistar} \trho:
\end{equation}
we fix $\phistar= G \Mstar/\rstar$, so $\tstar =\sqrt{\rstar^3/G
\Mstar}\equiv\tstarn$. The dMOND field equation~(\ref{eqdMOND}) in
dimensionless form writes
\begin{equation}
\tnabla\cdot( ||\tnabla\tphi||\tnabla\tphi)=4 \pi  {G \Mstar \az
  \over \phistar^2} \trho,
\end{equation}
so the natural choice is $\phistar=\sqrt{G \Mstar \az}$, and
$\tstar=\rstar(G\Mstar\az)^{-1/4}\equiv \tstard$. Finally, the MOND
field equation~(\ref{eqMOND}) in dimensionless form is
\begin{equation}
\tnabla\cdot\left[\mu(\kappa ||\tnabla\tphi||) \tnabla\tphi\right]=4
\pi {G \Mstar \over \rstar \phistar} \trho ,
\end{equation}
where $\kappa\equiv \phistar /\rstar\az$. In this case, as in the
Newtonian case, $\phistar=G \Mstar/\rstar$, so $\tstar=\tstarn$ and
$\kappa= G \Mstar /\rstar^2\az$.

\end{document}